\documentclass[12pt, draftclsnofoot, onecolumn]{IEEEtran} 
\usepackage{setspace}
\usepackage{clrscode}
\usepackage{amsmath}
\usepackage{amssymb}
\usepackage[dvips]{graphicx}
\usepackage{epsfig}
\usepackage{hyperref}
\usepackage{bm}
\usepackage{subfigure}
\usepackage{indentfirst}
\usepackage{float}
\usepackage{balance}
\usepackage{caption}
\usepackage{booktabs}
\newcommand{\figsize}{0.45}

\newtheorem{Rem}{Remark}

\begin{document}
\title{Space-Time Block Coding Based Beamforming for Beam Squint Compensation}
     \author{
\IEEEauthorblockN{Ximei Liu and Deli Qiao }
\thanks{This work has been supported in part by the National Natural Science Foundation of China (61671205) and the Shanghai Sailing Program (16YF1402600).}
\thanks{The authors are with the School of Information Science and Technology, East China Normal University, Shanghai, China 200241. Email: 52161214003@stu.ecnu.edu.cn, dlqiao@ce.ecnu.edu.cn.}}
\maketitle

\begin{abstract}
In this paper, the beam squint problem, which causes significant variations in radiated beam gain over frequencies in millimeter wave communication system, is investigated. A constant modulus beamformer design, which is formulated to maximize the expected average beam gain within the bandwidth with limited variation over frequencies within the bandwidth, is proposed. A semidefinite relaxation (SDR) method is developed to solve the optimization problem under the constant modulus constraints. Depending on the eigenvalues of the optimal solution, either direct beamforming or transmit diversity based beamforming is employed for data transmissions.  Through numerical results, the proposed transmission scheme can compensate for beam squint effectively and improve system throughput. Overall, a transmission scheme for beam squint compensation in wideband wireless communication systems is provided.
\end{abstract}

\section{Introduction}
The fifth generation (5G) communication is expected to provide up to hundreds of times higher date rate than fourth generation long term evolution (4G LTE) systems \cite{a1}. Millimeter wave (mmWave) band can provide the opportunity for GHz of spectrum and increase the data rate, and is viewed as a vital component of 5G \cite{a2}. However, the signals at mmWave frequencies experience significantly higher attenuation than signals below 6 GHz. Beamforming with a large number of antennas assumes a pivotal role in compensating the attenuation and maintaining a robust communication link \cite{a3}, \cite{a5}.

   In this paper, we consider analog beamforming with one antenna array, where each antenna is driven by a phase shifter and the antenna array is connected to a radio frequency (RF) chain. The design of phase shifters is challenging for ultra wideband communication systems.  The beam gain for any beam direction fluctuates with the frequency, i.e., \emph{beam squint}, which can limit the usable bandwidth in phased array antenna systems \cite{a6}, \cite{a7}.  Approaches to eliminate or reduce beam squint have been studied. In \cite{method1}, the authors have presented the implementation techniques based on true time delay. In \cite{method2}, to solve the beam squint problem in 60 GHz, the authors  have proposed three beamforming codebook designs. However, such approaches have these defects, such as high implementation cost, large size, or excessive power consumption, which make them unattractive to mobile wireless communication. In \cite{mmcai}, the authors have proposed a denser codebook design algorithm to compensate for beam squint. In addition, multiple noncontiguous band can be aggregated to increase the bandwidth, but at the same time, it introduces significant beam squint issues \cite{mmcai2}.

The space-time block coding (STBC) based beamforming has been proposed in \cite{STBC1}, \cite{STBC2} to achieve ominidirectional beampattern for massive multiple-input multiple-output (MIMO) systems to broadcast common information.
In this paper, we propose a STBC based beamforming to  guarantee little variation in beam gain for all subcarriers in the wideband system with a given \emph{beam focus}. The main contributions of the paper are three-fold.
\begin{itemize}
  \item We formulate the constant modulus beamforming problem that maximizes the average beam gain within the bandwidth with little variation in beam gain over frequencies within the bandwidth.
  \item We develop a semidefinite relaxation (SDR) technique based on the eigenvalue decomposition to obtain a suboptimal solution, and propose the STBC based beamforming scheme.
  \item Through numerical evaluations, we demonstrate the effectiveness of the proposed transmission scheme in compensating for the beam squint and enhancing the system throughput performance.
\end{itemize}

The rest of this paper is organized as follows. Section II  introduces the system model and the effect of beam squint. Section III proposes an effective beamforming scheme to minimize the beam squint for all subcarriers in the wideband system. Numerical results are given in Section IV, while Section V concludes this paper.
\section{Preliminaries}
\subsection{System Model}
\begin{figure}
\centering
\includegraphics[width=\figsize\textwidth,height=5cm]{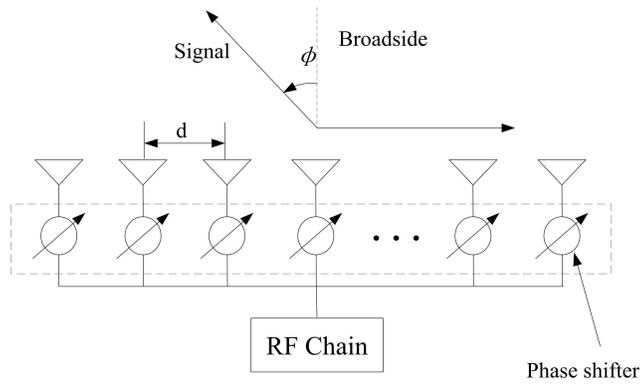}
\caption{Structure of a ULA with analog beamforming using phase shifters. }
\label{fig:ULAstructure}
\end{figure}

We consider a uniform linear array (ULA) antenna with $M$ radiation elements as shown in Fig. \ref{fig:ULAstructure} at the transmitter. The distance between any two adjacent antenna elements is denoted by $d$.
 The ULA downlink steering vector towards the virtual angle $\theta=\sin{(\phi)} $ with $\phi\in [-\frac{\pi}{2}, \frac{\pi}{2}]$ representing the angle-of-departure (AOD) has the form:
 \begin{small}
\begin{align}\label{eq:h}
\mathbf{a}(f, \theta)=[1, e^{j2\pi \frac{f}{c}d\theta}, \cdots, e^{j2\pi \frac{f}{c}(M-1)d\theta}]^{T},   \theta \in[-1,1],
\end{align}
\end{small}
where $[\cdot]^{T}$ indicates the transpose of a vector, $f$ is the signal frequency, and $c$ is the speed of light. The effect of phase shifters can be modeled by the beamforming vector:
\begin{align}
 \mathbf{w}=\frac{1}{\sqrt{M}}[e^{j\beta_1}, \cdots, e^{j\beta_m}, \cdots, e^{j\beta_M}]^T,
 \end{align}
 where $\beta_m$ is the phase of the $m$th phase shifter connected to the $m$th antenna in the array.

 The beam pattern in virtual angle $\theta$ is given by:
  \begin{align}
  p(f,\theta)=\mathbf{w}^{H}\mathbf{a}(f, \theta)\mathbf{a}(f, \theta)^{H}\mathbf{w}.
 \end{align}

    In \cite{mmcai}, to achieve the highest array gain for a beam focus $\theta_0$ among all possible beamforming vectors, the phase shifts should be:
 \begin{align}\label{eq:ps}
 \beta_m(\theta_0)=2\pi f_{c}c^{-1}(m-1)d\theta_0, m=1, 2, \cdots, M,
 \end{align}
    where $f_c$ is carrier frequency. The beam with phase shifts described in  (\ref{eq:ps}) is called  \emph{fine beam}. For a fixed array and carrier frequency, the array gains for a given $\theta$ are different for different frequencies, in particular, the beam gains of maximum frequency $f_{max}$ and minimum frequency $f_{min}$ for the beam focus $\theta_0$ are both significantly smaller than that of carrier frequency $f_{c}$, which is called beam squint.
\subsection{STBC Based Beamforming}
 To compensate for the beam squint and guarantee little variation in beam gain for all subcarriers in the wideband system, we consider a  Alamouti based beamforming scheme. The transmitted signals are first encoded using a STBC encoder. The parallel outputs of STBC encoder $s_i(t)$ and $s_i(t+1)$, $i=1, 2$, to  be transmitted in time slot $t$ and $t+1$ can be intuitively expressed in matrix as:
\begin{align}
\begin{bmatrix}
s_{1}(t)  &s_{1}(t+1)  \\
s_{2}(t)  &s_{2}(t+1)
\end{bmatrix}
=\begin{bmatrix}
\frac{1}{\sqrt{2}}s_{1}  &-\frac{1}{\sqrt{2}}s_{2}^{\ast}  \\
\frac{1}{\sqrt{2}}s_{2}  &\frac{1}{\sqrt{2}}s_{1}^{\ast}
\end{bmatrix},
\end{align}
where $\frac{1}{\sqrt{2}}$ is to ensure fixed total transmit power.

Subsequently, the STBC encoder outputs  $s_1$ and $s_2$  are processed by two transmit beamformers with weights $\mathbf{w}_1$ and $\mathbf{w}_2$, respectively, and summed together. Then, at the receiver side, through linear combination of the useful signals while ignoring the noise in two time slots, we have
\begin{small}
 \begin{align}
 \begin{bmatrix}
y_1(t)  \\
y_2^{\ast}(t+1)
\end{bmatrix}
 &=\frac{1}{\sqrt{2}}\begin{bmatrix}
\mathbf{w}_{1}^{H}\mathbf{a}(f, \theta)s_1+\mathbf{w}_{2}^{H}\mathbf{a}(f, \theta)s_2  \\
-\mathbf{a}(f, \theta)^{H}\mathbf{w}_{1}s_2+\mathbf{a}(f, \theta)^{H}\mathbf{w}_{2}s_1
\end{bmatrix}\nonumber\\
&=\frac{1}{\sqrt{2}}\begin{bmatrix}
\mathbf{w}_{1}^{H}\mathbf{a}(f, \theta) &\mathbf{w}_{2}^{H}\mathbf{a}(f, \theta)  \\
\mathbf{a}(f, \theta)^{H}\mathbf{w}_{2} &-\mathbf{a}(f, \theta)^{H}\mathbf{w}_{1}
\end{bmatrix}
\begin{bmatrix}
s_1  \\
s_2
\end{bmatrix}\nonumber\\
&=\mathbf{\Gamma}\begin{bmatrix}
s_1  \\
s_2
\end{bmatrix}.
 \end{align}
 \end{small}
 Then, with weights $\mathbf{\Gamma}^{H}$, we can obtain the equivalent processing gain for the transmitted signals $s_1$ and $s_2$ as
  \begin{align}
  \begin{bmatrix}
p_1  \\
p_2
\end{bmatrix}
=\mathbf{\Gamma}^{H}\mathbf{\Gamma}
=\frac{1}{2}\begin{bmatrix}
\alpha_1^2+\alpha_2^2  & 0\\
0     & \alpha_1^2+\alpha_2^2
\end{bmatrix},
 \end{align}
 where $\alpha_{1}^{2}=|\mathbf{w}_{1}^{H}\mathbf{a}(f, \theta)|^2$ and $\alpha_{2}^{2}=|\mathbf{w}_{2}^{H}\mathbf{a}(f, \theta)|^2$. That is, it is possible for us to expressed the virtual beam pattern for the transmitted signal in virtual angle $\theta$ as
\begin{align}\label{eq:beamgain2}
  p(f,\theta)=\frac{1}{2}(|\mathbf{w}_1^{H}\mathbf{a}(f, \theta)|^2+|\mathbf{w}_2^{H}\mathbf{a}(f, \theta)|^2).
 \end{align}
 \subsection{Hybrid Beamforming}
\begin{figure}
\centering
\includegraphics[width=\figsize\textwidth,height=4cm]{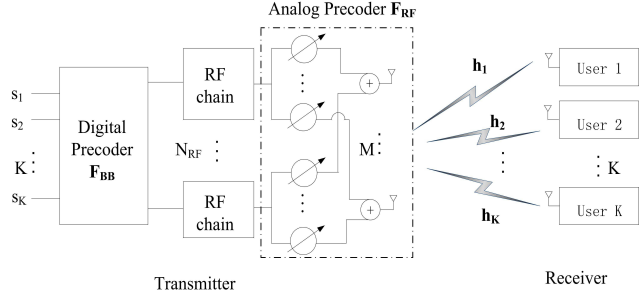}
\caption{Block diagram of an MU-MIMO  hybrid beamfoming communication system with single-antenna users.}
\label{fig:hybridmodel}
\end{figure}

  To achieve the best possible compromise in terms of spectral efficiency and quality of multistream separation, hybrid analog-digital beamforming is applied in mmWave communication systems. Consider an multi-users multiple-input multiple-output (MU-MIMO) communication system with single-antenna users shown in Fig. \ref{fig:hybridmodel} for example. The transmitter equipped with $M$ antennas and $N_{RF}$ RF chains transmits $K$ data streams to $K$ single-antenna users. In the fully-connected hybrid beamforming structure, the transmitter is assumed to apply a baseband precoder $\mathbf{F}_{BB}\in\mathbb{C}^{N_{RF}\times K}$ followed by a RF precoder $\mathbf{F}_{RF}\in\mathbb{C}^{M\times N_{RF}}$. Then the received signals of the users are given by
 \begin{align}
 \mathbf{y}=\mathbf{H}^{H}\mathbf{F}_{RF}\mathbf{F}_{BB}\mathbf{s}+\mathbf{n},
 \end{align}
 where $\mathbf{s}=[s_1, s_2, \cdots, s_K]^T\sim\mathcal{CN}(\mathbf{0},\mathbf{I}_{K})$ denotes the signal vector, $\mathbf{H}=[\mathbf{h}_1, \mathbf{h}_2, \cdots, \mathbf{h}_K]\in\mathbb{C}^{M\times K}$ is composed of the the downlink channel vectors for the users, and $\mathbf{n}\in \mathbb{C}^{K\times 1}$ with $n_k\sim \mathcal{N}(0, N_0)$ is the Gaussian noise corrupting the received signals. Thus, beam squint still affects the hybrid beamforming system as in the analog beamforming system for each RF chain related analog beamforming vector. Therefore, we would like to note that the method proposed in this paper can still be applied to the hybrid beamforming systems, albeit the target beam pattern may vary. Henceforth, we focus on analog beamforming in this paper for simplicity.
\section{Beamforming Design}
Suppose the bandwidth of the signal is B. Let $\mathcal{D}=[f_c-f_0, f_c+f_0]=\mathcal{D}_{ML}\cup \mathcal{D}_{SL}$ be the total frequency range, where $\mathcal{D}_{ML}=[f_c-\frac{B}{2},f_c+\frac{B}{2}]$ and $\mathcal{D}_{SL}=\mathcal{D}\backslash\mathcal{D}_{ML}$ represent the frequency range for the main lobe and the side lobe, respectively, with $f_0\gg B$.

The beam pattern with weight $\mathbf{w}$ over different frequencies with given beam focus $\theta_0$ can be expressed as:
 \begin{align}\label{eq:p}
p(f, \theta_0)=\mathbf{w}^{H}\mathbf{a}(f, \theta_0)\mathbf{a}(f, \theta_0)^{H}\mathbf{w}, f\in\mathcal{D}. \nonumber
\end{align}
 We would like little variation in the beam pattern for different frequencies within the bandwidth. The average beam gain $P_{I}$  in the beam covering $\mathcal{D}_{ML}$ is given by:
\begin{align}
P_{I}=\frac{1}{B}\int_{f\in\mathcal{D}_{ML}}\mathbf{w}^{H}\mathbf{a}(f, \theta_0)\mathbf{a}(f, \theta_0)^{H}\mathbf{w} \, df.
\end{align}
 Meanwhile, we want the average beam gain in the side lobe:
\begin{align}
P_{o}=\frac{1}{2f_0-B}\int_{f\in \mathcal{D}_{SL}}\mathbf{w}^{H}\mathbf{a}(f, \theta_0)\mathbf{a}(f, \theta_0)^{H}\mathbf{w} \, df
\end{align}
to be limited for large power efficiency.

 Therefore, the design of $\mathbf{w}_i$ can be formulated as the following optimization problem\footnote{Note that the desired beam pattern may vary for hybrid beamforming, which can be approximated by discrete samplings of $\theta$ for (\ref{eq:1A})-(\ref{(eq:1D)}) \cite{dqiaobeam}. The same method in the following still applies.}:
\begin{small}
\begin{subequations} \label{eq:1}
 \begin{align}
 \textbf{P1:}&\nonumber\\
&\max_{\mathbf{w}} \quad P_{I} \label{eq:1A}\\
& \text{s.t.}  \quad \gamma\leq \mathbf{w}^{H}\mathbf{a}(f, \theta_0)\mathbf{a}(f, \theta_0)^{H}\mathbf{w}\leq 10^{\frac{\varepsilon}{10}}\gamma,  f\in\mathcal{D}_{ML},                      \label{(eq:1B)}\\
  &\hspace{.7cm}\mathbf{w}^{H}\mathbf{a}(f, \theta_0)\mathbf{a}(f, \theta_0)^{H}\mathbf{w}\geq 0,  f\in\mathcal{D}_{SL},  \label{(eq:1C)}\\
  &\hspace{.7cm}P_{o}\leq \gamma \zeta,       \label{(eq:1D)}\\
  & \hspace{.7cm}w_{j}^{\ast}w_{j}=\frac{1}{M}, j=1,\cdots, M,                                             \label{eq:1E}
  \end{align}
 \end{subequations}
 \end{small}
where $\varepsilon$ indicates the strength of the ripples within the beam range in dB and $\gamma$ is some value satisfying (\ref{eq:1E}) and $\zeta$ is a small adjustable restriction parameter.
\begin{Rem}
Note that the constraint (\ref{(eq:1C)}) is to ensure that the beam gain with the frequencies outside the bandwidth is non-negative, which is presented here for the convenience of the following transformation of the problem. Without this constraint, the following transformed optimization problem (\ref{eq:2}) may return incorrect solution.
\end{Rem}

It is observed that both the objective function and constraints in (\ref{eq:1})  are linear in the matrix $\mathbf{w}\mathbf{w}^{H}$. By introducing a variable $\mathbf{X}=\mathbf{w}\mathbf{w}^{H}$, it is clear that
\begin{align}
\mathbf{X}\succeq \mathbf{0}, \text{rank}(\mathbf{X})= 1.
\end{align}
$\mathbf{X}\succeq \mathbf{0}$ means that matrix $\mathbf{X}$ is positive semidefinite, 
However, the rank constraint is nonconvex. Dropping the rank constraint, the standard SDR for problem (\ref{eq:1}) becomes \cite{sdp}:
 \begin{small}
 \begin{subequations} \label{eq:2}
  \begin{align}
   \textbf{P2:}&\nonumber\\
   &\max_{\mathbf{X}} \quad\frac{1}{B}\int _{f \in \mathcal{D}_{ML}}\text{tr}(\mathbf{a}(f, \theta_0)\mathbf{a}(f, \theta_0)^{H}\mathbf{X}) df  \label{eq:2A}\\
  &\text{s.t.} \quad \gamma\leq \text{tr}(\mathbf{a}(f, \theta_0)\mathbf{a}(f, \theta_0)^{H}\mathbf{X})\leq \gamma 10^{\frac{\varepsilon}{10}}, f\in\mathcal{D}_{ML},                      \label{(eq:2B)}\\
  &\hspace{.7cm}\text{tr}(\mathbf{a}(f, \theta_0)\mathbf{a}(f, \theta_0)^{H}\mathbf{X})\geq 0, f\in\mathcal{D}_{SL},  \label{(eq:2C)}\\
  &\hspace{.7cm}\frac{1}{2f_0-B}\int _{f\in\mathcal{D}_{SL}} \text{tr}(\mathbf{a}(f, \theta_0)\mathbf{a}(f, \theta_0)^{H}\mathbf{X}) df \leq\gamma\zeta,       \label{(eq:2D)}\\
  &\hspace{.7cm}\mathbf{X}(j,j)=\frac{1}{M}, j=1,\cdots, M,  \label{(eq:2E)}\\
  &\hspace{.7cm}\mathbf{X}\succeq \mathbf{0}.    \label{(eq:2F)}
  \end{align}
  \end{subequations}
  \end{small}
Clearly, the problem $\textbf{P2}$ falls into the category of convex programming. To solve it, the package CVX for specifying and solving convex programs is used \cite{CVX}. In general, there is no guarantee that $\textbf{P2}$ always admits a rank-one or rank two solution. In such a case, eigenvalue decomposition and projection are needed to extract a feasible solution for $\textbf{P1}$.
Let $r$ be the number of eigenvalues greater than some small value $\xi>0$. We have the approximate eigen-decomposition as
\begin{align}
 \mathbf{X}^{\star}\approx\sum_{i=1}^{r}\mu_i\mathbf{u}_i\mathbf{u}_i^{H},
 \end{align}
 where $\mu_1\geq \mu_{2}\geq\cdots\geq \mu_{r}>\xi$ are the eigenvalues, and $\mathbf{u}_1, \cdots, \mathbf{u}_r$ are the respective eigenvectors. Note that $\text{tr}(\mathbf{X}^{\star})=1$, so $\sum_{i=1}^{M}\mu_i=1$. If $r=1$, we output $\mathbf{w}=\frac{1}{\sqrt{M}}\exp(j\angle{\mathbf{u}}_1)$; if $r>1$, the STBC based beamforming scheme will be incorporated, and we extract two feasible solutions $(\mathbf{w}_1, \mathbf{w}_2)$ from the $r$ effective eigenvectors that maximize the virtual beam pattern. The detailed computational procedure is shown by Algorithm 1. Note that the algorithm can be performed offline and the complexity is not a big issue.
\begin{table}
\begin{center}
 \begin{tabular}{lcl}
 \toprule
 \textbf{Algorithm 1. Generation of Beamforming Vectors.}   \\
 \midrule
 1: \textbf{Input} $\xi$, $\theta_0$;\\
 2: Solve problem $\textbf{P2}$ to obtain $\mathbf{X}^{\star}$;\\
 3: Perform eigendecomposition of $\mathbf{X}^{\star}$, obtain the eigenvalues \\
 $\mu_1\geq \mu_{2}\geq\cdots\geq \mu_{M}$ and the respective eigenvectors $\mathbf{u}_1, \mathbf{u}_2,$\\$\cdots, \mathbf{u}_M$;\\
 4: Determine $r=\arg_{i}\{\mu_i >\xi\, \& \,\mu_{i+1}\leq\xi\}$;\\
 5:  \textbf{If} r==1     \\
 6:  \hspace{0.7cm} $\mathbf{w}=\frac{1}{\sqrt{M}}\exp(j\angle\mathbf{u}_1)$;\\
 7:      \textbf{else}\\
 8:         \hspace{0.7cm}     $(\mathbf{w}_1, \mathbf{w}_2)=\{(\frac{1}{\sqrt{M}}\exp(j\angle\mathbf{u}_i), \frac{1}{\sqrt{M}}\exp(j\angle\mathbf{u}_{j})),$ \\
          \hspace{3.2cm}     $i, j\in\{1, \cdots, r\}, i\neq j\}$, \\
    \hspace{0.5cm}  find  $(\mathbf{w}_1, \mathbf{w}_2)$ that maximizes the virtual beam gain  (8);  \\
 9:  \textbf{End}\\
 10: \textbf{Output} $\mathbf{w}$ or $(\mathbf{w}_1, \mathbf{w}_2)$.\\
 \bottomrule
 \end{tabular}
 \end{center}
 \end{table}
\section{Numerical Results}
In this section, we evaluate the performance of the proposed transmission scheme and consider the fine beam as the baseline \cite {mmcai}. Due to the symmetry of each beam pattern, we only show the performance of half the beam focus range $\theta_0 \in (0,1)$.
 We assume that the transmitter is equipped with $M=$64 ULA antennas with wavelength $d=\lambda_c/2$ spacing between adjacent antennas. $\xi$ is set to be 0.1.

\begin{figure}
\centering
\includegraphics[width=\figsize\textwidth]{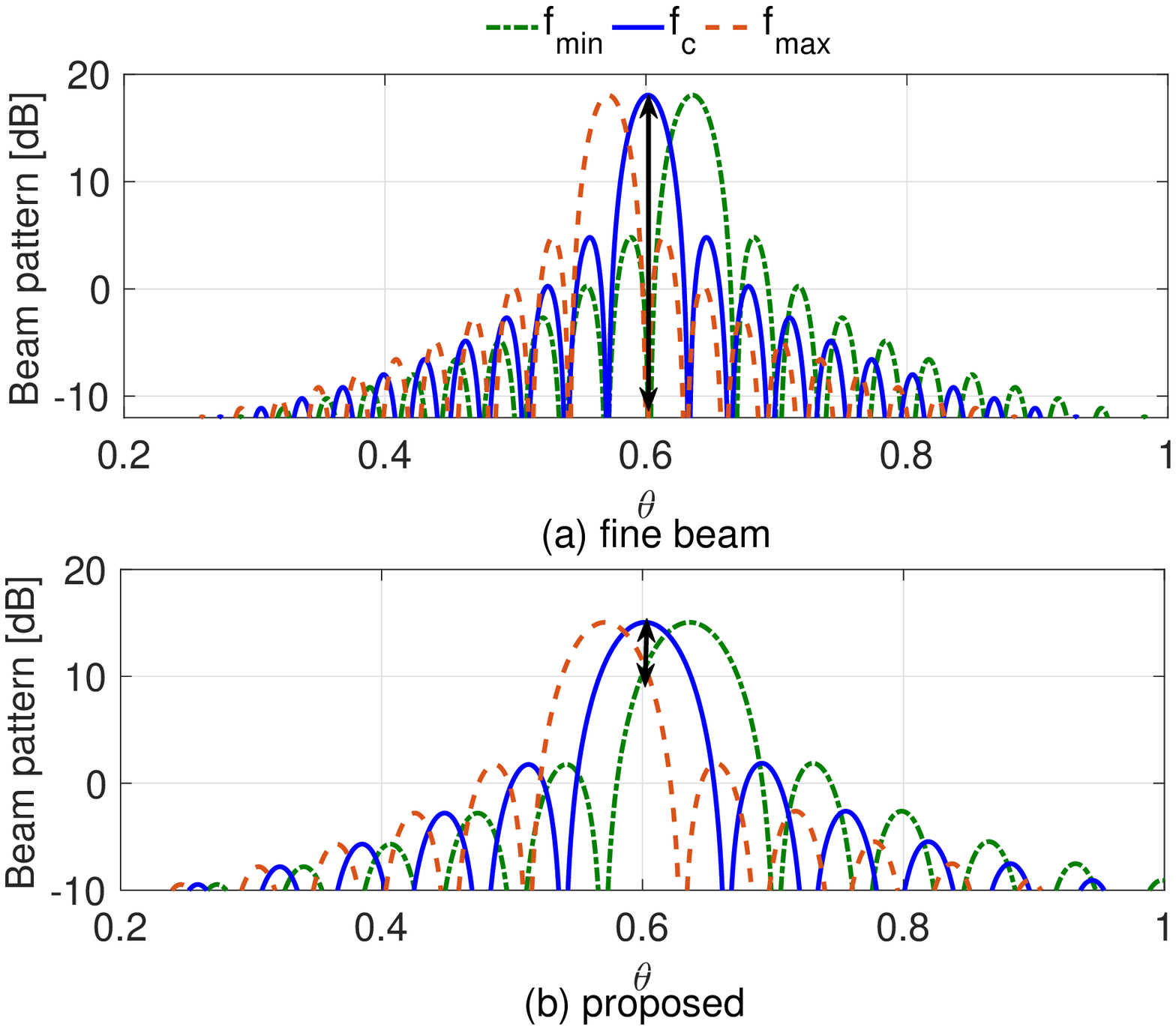}
\caption{The comparison of beam gain for proposed scheme and fine beam in a wideband system with beam focus $\theta_0=$0.6, carrier frequency $f_c=$28 GHz,  maximum frequency $f_{max}=$29.5 GHz, and minimum frequency $f_{min}=$26.5 GHz.}
\label{fig:beampattern2}
\end{figure}

  Fig. \ref{fig:beampattern2} illustrates the comparison of beam gain between proposed scheme and fine beam in a wideband system with beam focus $\theta_0=0.6$, carrier frequency $f_c=28$ GHz,  maximum frequency $f_{max}=29.5$ GHz, and minimum frequency $f_{min}=26.5$ GHz. The beam gain degradation with fine beam for $f_{max}$ or $f_{min}$ is much larger, about 31 dB, which is shown by the double-arrow in Fig. \ref{fig:beampattern2}(a). On the other hand, the beam gain degradation with the proposed beam for $f_{max}$ or $f_{min}$ is only around 5 dB, which is shown by the double-arrow in Fig. \ref{fig:beampattern2}(b). Therefore, the proposed scheme can well compensate for the beam squint in  wideband communication systems.
\begin{figure}
\centering
\includegraphics[width=\figsize\textwidth]{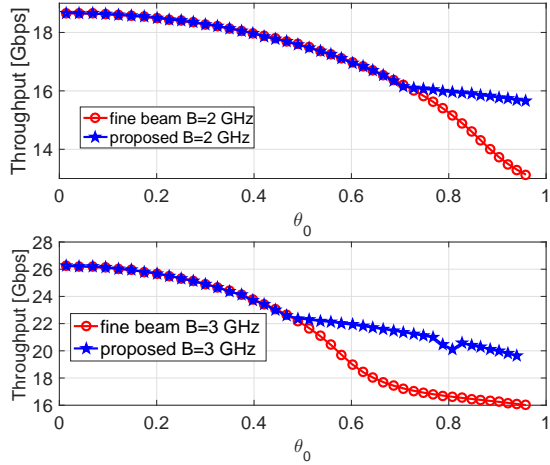}
\caption{Throughput of the receiver at the different beam focus.}
\label{fig:throughput}
\end{figure}

The problem of beam squint caused by wideband beamforming will lead to the decrease of system throughput. Assume line-of-sight (LoS) channel between the transmitter node and  receiver, the channel will be in the form of (\ref{eq:h}). We assume 0.2 mW total transmit power and -74 dBm/Hz background noise in the wireless communication system. Fig. \ref{fig:throughput} shows the system throughput for the transmission bandwidth of 2 GHz and 3 GHz with 28 GHz carrier frequency. From the simulation results, it can be seen that the throughput used the proposed scheme has an obvious growth over that of fine beam. For the 2 GHz bandwidth system, the throughput is improved when the beam focus $\theta_0$ is greater than 0.7, and the maximum improvement is up to 19$\%$. For the 3 GHz bandwidth system, the throughput is improved when the beam focus $\theta_0$ is greater than 0.49, and the maximum improvement is up to 24.6$\%$. The system throughputs of the proposed scheme and fine beam are equal when direct beam transmission structure is adopted. In a word, more enhancement of the system performance caused by the beam squint compensation can be expected in ultra wideband communication system.
\section{Conclusion}
In this paper, we have investigated the beam squint problem in millimeter wave communication system. We have proposed a beamforming design that maximizes the average beam gain within the bandwidth with little fluctuation in different frequencies  within the bandwidth while minimizing the average beam gain outside the bandwidth, and developed a STBC based beamforming scheme.  Through numerical results, we have demonstrated that our proposed transmission scheme can effectively compensate for the beam squint and improve the throughput performance for wideband communication system in certain cases.

\balance

\end{document}